\documentclass[12pt]{article}
\usepackage{graphicx,amsmath,amsfonts}
\newcommand\ba{\begin{eqnarray}}
\newcommand\ea{\end{eqnarray}}
\newcommand\dd{d}
\date{}

\begin{document}
\title{Charge asymmetry in $\pi^+\pi^-$ electroproduction on proton at high energies
as a test of $\sigma,\rho$ mesons degeneration}

\author{A.I.Ahmedov$^{1,2}$, V.V.Bytev$^{1}$ and E.A.Kuraev$^{1}$}
\maketitle
\begin{center}
$^{1}$ {\small Joint Institute for Nuclear Research, Dubna, Russia}, \\
$^{2}$ {\small Institute of Physics, Azerbaijan National Academy \\
of Sciences, Baku}
\end{center}
\noindent
{\bf Abstract} \\
The charge asymmetry induced by interference of amplitudes with $\sigma$ and
$\rho$ mesons decaying to $\pi^+\pi^-$ pair created in the fragmentation region
of proton, suggest to be a test of degeneration hypothesis of $\rho$ and
$\sigma$ mesons. Some numerical estimations are given.
\section{Introduction}

Theoretical reasons for existence of scalar neutral meson ($\sigma$-meson)
was formulated in late 60-th \cite{Salam}. It was recognized that the chiral as well
as scaling symmetry of strong interactions are violated, which can be
realized within effective lagrangian by including scalar field $\sigma(x)$.
In the frame QCD it was shown that breaking of scale invariance is related to
the trace of energy-momentum tensor \cite{Adler}.
In the papers of Schehter, Ellis and Lanik \cite{Ellis} the effective QCD lagrangian with
broken scale and chiral symmetries was constructed,where scalar gluonic
current was related with $\sigma(x)$: $G_{\mu\nu}^2\sim
m_\sigma^4\sigma(x)^4/G_0,G_0=<0|(\alpha_s/\pi G_{\mu\nu}^2|0>=0.017GeV^4$-
is gluonic condensate. Besides the widths of 2 pion and 2 gamma decay
channels it was obtained:
\ba
\Gamma(\sigma\to \pi^+\pi^-)=\frac{m_\sigma^5}{48\pi G_0}, \quad
\Gamma(\sigma\to\gamma\gamma)=\frac{3}{4}\biggl(\frac{R\,\alpha}{8\pi^2}\biggr)^2\Gamma(\sigma\to \pi^+\pi^-),
\ea
with $R=\sigma(e^+e^-\to hadrons)/\sigma(e^+e^-\to \mu^+\mu^-)$.
Experimental evidence of possible existence of $\sigma(750)$ was obtained
in CERN experiments for process $\pi^-p\to\pi^-\pi^+n$ \cite{Svec, Alex} with polarized
target. In S. Weinberg paper \cite{Weinberg}  it was shown the relation
$m_\rho=m_\sigma$ as a consequence of broken chiral symmetry. A similar
statement follows from the analysis of superconvergent sum rules for
helicity amplitudes, as was shown in the paper of Gillman and Harari \cite{Gilman}.
Really, taking this value for $\sigma$ meson we obtain for its total
(mainly 2 pion) width $\Gamma(\sigma\to 2\pi)=150 MeV=\Gamma_\rho$.

To obtain the independent evidence of $\sigma$ and besides the validity of
the degeneracy $m_\rho=m_\sigma$, $\Gamma_\rho=\Gamma_\sigma$ we suggest to
measure the charge asymmetry of two pion production at electron-proton
collisions:
\ba
\label{kinem}
e(p_1)+p(p)\to e'(p_1')+p'(p')+\pi^+(q_+)+\pi^-(q_-),
\ea
which is defined as follows:
\ba
A_c&=&\frac{d\sigma(q_1,q_2)-d\sigma(q_2,q_1)}
{d\sigma(q_1,q_2)+d\sigma(q_2,q_1)} \nonumber \\
&=&\frac{N(\pi^+(q_1),\pi^-(q_2))-N(\pi^+(q_2),\pi^-(q_1))}
{N(\pi^+(q_1),\pi^-(q_2))+N(\pi^+(q_2),\pi^-(q_1)) },
\ea
where $d\sigma(q_1,q_2)$ means the inclusive cross section with
$\pi^-$ meson with momentum $q_1$ and $\pi^+$ with momentum $q_2$,
$N(\pi^+(q_1),\pi^-(q_2))$-number of corresponding events.

\section{Calculation of asymmetry}

Charge asymmetry can be more pronounced at invariant mass of pions close
to the $\rho$ meson mass,where due to Breit-Wigner enhancement of cross
section the counting rate is expected to be large.

\begin{figure}[htbp]
\begin{center}
\includegraphics[scale=1]{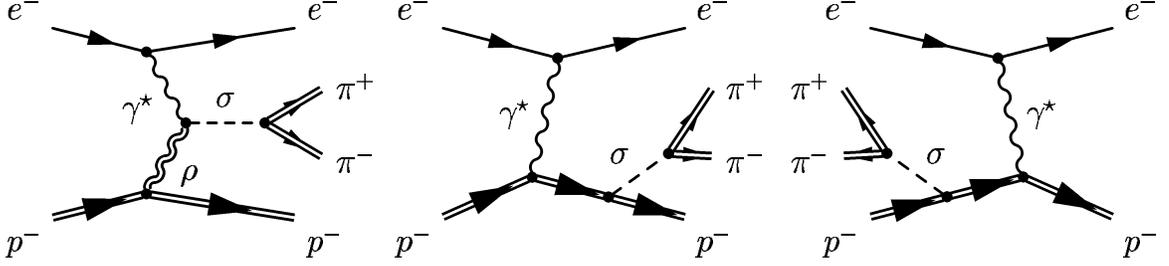}
\end{center}
\caption{Feynman Diagrams (FD) for matrix element ${\cal M}_\sigma$. By exchange $\sigma\to\rho, \rho\to\sigma$
one can obtain FD for ${\cal M}_\rho$ }
\label{fdiag}
\end{figure}

Matrix element in this
region can be put in form $\cal M=M_\rho+M_\sigma$ (see fig.\ref{fdiag}). Then the charge asymmetry
will have a form
\ba
A_c= \frac{2{\cal M}_\rho({\cal M}_\sigma)^*}{|{\cal M}_\sigma|^2+|{\cal M}_\rho|^2}.
\ea
To obtain the realistic estimation magnitude of the effect we take into
account only Feynman amplitudes containing the $\sigma,\rho$ intermediate
states.

In Appendix we give the cross section and the asymmetry in kinematics of
fragmentation of proton both in exclusive and inclusive set-up.

Matrix elements have the form:
\begin{gather}
{\cal M}_\sigma=J_\mu \bar{u}(p')T^\mu u(p)R_\sigma g_{\sigma \pi \pi}g_{\sigma nn}, \\ \nonumber
{\cal M}_\rho=J_\mu \bar{u}(p')T^{\mu\nu}u(p)(q_--q_+)^\nu R_\rho g_{\rho \pi\pi}
g_{\rho n n},
\end{gather}
with (for kinematics  see (\ref{kinem}))
\begin{gather}
J_\mu=\frac{4\pi\alpha}{Q^2}\bar{u}(p_1')\gamma_\mu u(p_1),\quad
R_{\sigma,\rho}=\frac{1}{s_1-m_{\sigma,\rho}^2+i(m\Gamma)_{\sigma,\rho}},\\ \nonumber
Q^2=-q^2, \quad s_1=q_2^2, \quad q_2=q_++q_-,\quad q=p_1-p_1'.
\end{gather}
Besides
$g_{\rho \pi \pi}=4\sqrt{3\pi/\beta_{\rho}^3}\sqrt{\Gamma_\rho/m_\rho}$ and
$ g_{\sigma \pi \pi}=4\sqrt{\pi/\beta_{\sigma}}\sqrt{m_\sigma\Gamma_\sigma}$ -
are the coupling
constant, $\beta_{\rho,\, \sigma}=\sqrt{1-(4m_\pi^2/m^2_{\rho,\,\sigma})}$ - the
velocity of pions in the rest frame of decaying $\sigma$ or $\rho$ mesons.

The hadronic currents obey the current conservation condition:
\ba
\bar{u}(p')T_\mu u(p)q^\mu=0, \quad \bar{u}(p')T^{\mu\nu} u(p)q_\mu=0, \quad
\bar{u}(p')T^{\mu\nu} u(p) q_{2\nu}=0.
\ea
The expressions for $T$ we had used are:
\ba
T^\mu=\Lambda
\frac{1}{q_1^2-m_\rho^2}R^{\mu\nu}_1\gamma_\nu+\frac{\hat{p}+\hat{q}+m_p}{(p+q)^2-m_p^2}\gamma^\mu+
\gamma_\mu\frac{\hat{p}'-\hat{q}+m_p}{(p'-q)^2-m_p^2},
\ea
and
\ba
T^{\mu\nu}=\frac{1}{\Lambda}\frac{1}{q_1^2-m_\sigma^2}R^{\mu\nu}_2+\gamma_\nu
\frac{\hat{p}+\hat{q}+m_p}{(p+q)^2-m_p^2}\gamma_\mu+\gamma_\mu
\frac{\hat{p}'-\hat{q}+m_p}{(p'-q)^2-m_p^2}\gamma_\nu,
\ea
with $\Lambda=g_{\rho nn}/g_{\sigma nn}$, $q_1=p-p'$.
The vertex $\gamma^*\sigma\rho$ we parameterize as
\ba
R^{\mu\nu}_{1,2}=\frac{g \, m}{m^2+Q^2}(q^\nu q_{1,2}^\mu-g^{\mu\nu} qq_{1,2}),
\ea
which is an anzatz, inspired by low-order triangle Feynman diagram
calculation, $m=300 MeV$ is the constituent quark mass. The coupling $g$ is
chosen in such a way to reproduce the $(g_{\rho \pi
\pi}/e)^2\Gamma(\sigma\to \gamma\gamma)$. The factor $(g_{\rho \pi
\pi}/e)^2$ is introduced to take into account the replacement of one of
photons by vector meson. Our estimate gives $g\approx 2$.

By calculating matrix element which are given above we obtain
charge asymmetry in the form:
\ba
\label{Ac}
A_c=\frac{2[(s_1-m_\rho^2)(s_1-m_\sigma^2)+m_\rho
m_\sigma\Gamma_\rho\Gamma_\sigma]Q^{\mu\lambda}(q_--q_+)^\nu
Z_{\mu\nu\lambda}\Lambda g_{\rho \pi \pi}g_{\sigma \pi \pi}}{Q^{\eta\sigma}[|R_\rho|^{-2}Z_{\eta\sigma}g^2_{\sigma
\pi \pi}+\Lambda^2|R_\sigma|^{-2}Z_{\eta\mu_1\sigma\nu_1}g^2_{\rho \pi
\pi}(q_--q_+)_{\mu_1}(q_--q_+)_{\nu_1}]}
\ea
with
\ba
Q^{\mu\nu}&=&2p_1^\mu p_1^\nu+(Q^2/2)g^{\mu\nu},\\ \nonumber
Z_{\mu\nu\lambda}&=&Tr(\hat{p}'+m_p)T_\mu(\hat{p}+m_p)\tilde{T}_{\lambda\nu}, \\ \nonumber
Z_{\eta\sigma}&=&Tr(\hat{p}'+m_p)T_\eta(\hat{p}+m_p)\tilde{T}_\sigma, \\ \nonumber
Z_{\eta\mu_1\sigma\nu_1}&=&Tr(\hat{p}'+m_p)T_{\eta\mu_1}(\hat{p}+m_p)\tilde{T}_{\sigma\nu_1}.
\ea
For inclusive set-up (pions as well as the scattered electron are
detected) the numerator and the denominator must be integrated by phase
volume of the scattered proton.

\section{Discussion}
\label{Disc}

We neglect above final-state pion and pion-nucleon interaction which cause
the pion-pion scattering phases \cite{Teryaev}. It can be justified within
5\% accuracy for the case of rather large effective mass of pions in the
final state.

In the Tables 1-4 we present $x_{+},x_{-}$ distribution (which means energy fraction of
$\pi^+$ and $\pi^-$ in the case of the energies of experiment H1) for the equal
mass case $m_{\sigma} =m_{\rho} =769$ MeV and different mass case.

The integration of asymmetry was made over variables $\vec{q}_\pm$  in the region $0.01-0.9$ GeV
for each $x,y$ component of the vector  $\vec{q}_\pm$:
\begin{gather}
Ac_{int}(x_+,x_-,Q)=\frac{\int\dd^2q_+\int\dd^2q_-
2{\cal M}_\rho({\cal M}_\sigma)^*}
{\int\dd^2q_+\int\dd^2q_-
(|{\cal M}_\sigma|^2+|{\cal M}_\rho|^2)}
\end{gather}
We see that the case $m_{\sigma} =m_{\rho}$ can be unambiguously
separated from the $m_{\sigma} \neq m_{\rho}$ case within 5\% accuracy
of experimental data.

Asymmetry have a magnitude of order 1 $(|A_c| \sim 1)$ in the fragmentation region
of initial proton for small values of momentum transferred $Q \sim 0.5$ GeV $\ll \sqrt s \sim 100$ GeV
(DESY, H1). At higher $Q$ it degrees rapidly: for $Q \sim 3$ GeV $|A_c|<0.1$.

In tables 5-7 we give asymmetry as function of angles $\theta_+,\theta_-$ for HERMES case
of energies, $\theta_\pm=\widehat{p_eq_\pm}$, integrated over $\varepsilon_+,\varepsilon_-$
(energies of $\pi_+,\pi_-$) in the region $0.5-20$ GeV, and over angles $\phi_+,\phi_-$
($\phi_\pm=\widehat{q_{\bot}q_{\pm\bot}}$) in the region $0-2\pi$, $Q=0.2$ GeV:
\begin{gather}
Ac_{int}(\theta_+,\theta_-,Q=0.2)=\frac{\int\limits_{0}^{2\pi}\int\limits_{0}^{2\pi}\dd\phi_+\dd\phi_-
\int\int\dd\varepsilon_+\dd\varepsilon_-
2{\cal M}_\rho({\cal M}_\sigma)^*}
{\int\limits_{0}^{2\pi}\int\limits_{0}^{2\pi}\dd\phi_+\dd\phi_-
\int\int\dd\varepsilon_+\dd\varepsilon_-
(|{\cal M}_\sigma|^2+|{\cal M}_\rho|^2)}
\end{gather}
Note that asymmetry in tables 5-7 turns to zero not only when $\theta_+=\theta_-$, but also at points
$\theta_+=-\theta_-$.

In all numerical estimates given in tables we used the next values:
\begin{gather}
\Lambda=\frac{g_{\rho n n}}{g_{\sigma n n }}=1, \quad  m=0.300\,\, GeV,\quad m_\rho=0.769\,\,GeV,\\ \nonumber
\Gamma_\rho=\Gamma_\sigma=150 \,\, GeV,\quad M_p=0.98 \,\, GeV, \quad m_\pi=0.139 \,\, GeV.
\end{gather}

\section{Acknowledgements}
One of us (V.V.B) is grateful to HERMES Collaboration for valuable
discussion and warm hospitality during May 2003.
We are grateful to grant RFFI grant 03-02-17077 and two of us E.K. and V.B.
to grant INTAS 00-00-366. \\
E.A.K. is grateful to L.Shimanovski and O.Teryaev for critical discussions.

\section{Appendix}

In proton fragmentation region (invariant mass of the scattered proton jet)
$(p'+q_+ +q_-)^2 \ll s =2pp_1$ we can use Sudakov decomposition of 4-momenta (\ref{kinem}):
\begin{gather}
\label{g1}
q=\beta\tilde{p}_1+\alpha\tilde{p}+q_\bot,\quad
q_\pm=\beta_\pm\tilde{p}_1+x_\pm\tilde{p}+q_{\pm\bot},\quad
p'=\beta' \tilde{p}_1+x\tilde{p}+p_\bot, \\ \nonumber
\tilde{p}_1 =p_1 -p\frac{m_1^2}{s},\quad \tilde{p} =p-p_1 \frac{m_p^2}{s},
\end{gather}
with conservation law and on mass shell conditions:
\begin{gather}
\label{g2}
\frac{m_p^2}{s}+\beta=\beta_++\beta_-+\beta',\quad \vec{q}=\vec{q}_++\vec{q}_-+\vec{p},\quad
x+x_-+x_+=1, \\ \nonumber
s\beta_\pm=A_\pm=\frac{1}{x_\pm}[m_\pi^2+\vec{q}^2_\pm],\quad
s\beta'=A=\frac{1}{x}[m_p^2+\vec{p}^2].
\end{gather}
Above we put designations for transversal components of 4-vectors:
\ba
\label{g3}
\vec{q}_\pm^2=-q_{\pm\bot}^2, \quad \vec{p}^2=-p_\bot^2,
\ea
and $m_p$ $(m_\pi)$ are mass of proton (pion).

Phase space volume
$$d\Gamma=(2\pi)^{-8}\frac{d^3p_1'}{2\epsilon_1'}\frac{d^3p'}{2\epsilon'}
\frac{d^3q_+}{2\epsilon_+}\frac{d^3q_-}{2\epsilon_-}\delta^4(p_1+p-p'-p_1'-q_+-q_-)$$
can be transformed to the form:
\ba
d\Gamma=(2\pi)^{-8}\frac{d^2\vec{q}d^2\vec{q}_+d^2\vec{q}_-}{8s}
\frac{dx_+dx_-}{x_+x_-(1-x_+-x_-)}.
\ea
Matrix element,using the Gribov's representation for metric tensor
$$g^{\mu\nu}=g^{\mu\nu}_\bot+\frac{2}{s}[\tilde{p}_1^\mu\tilde{p}^\nu+
\tilde{p}_1^\nu\tilde{p}^\mu]$$
can be written in form:
\begin{gather}
{\cal M}=\frac{8\pi\alpha i s}{q^2}N(\Phi_\rho
g_{\rho \pi \pi}g_{\rho nn}+\Phi_\sigma g_{\sigma \pi \pi}g_{\sigma nn}),\\ \nonumber
N=\frac{1}{s}\bar{u}(p_1')\tilde{p} u(p_1),\quad \Sigma|N|^2=2;
\end{gather}
and
\ba
\Phi_{\sigma,\rho}=\frac{R_{\sigma,\rho}}{s}\bar{u}(p')O_{\sigma,\rho}u(p),
\ea
with
\begin{gather}
O_\sigma=as\hat{q}+b\hat{p}_1+\frac{1}{d}(\hat{p}+\hat{q}+M)\hat{p}_1+
\frac{1}{d'}\hat{p}_1(\hat{p}'-\hat{q}+M), \\
O_\rho=cs+\frac{1}{d}(\hat{q}_--\hat{q}_+)(\hat{p}+\hat{q}+M)\hat{p}_1+
\frac{1}{d'}\hat{p}_1(\hat{p}'-\hat{q}+M)(\hat{q}_--\hat{q}_+).
\end{gather}
Asymmetry (\ref{Ac}) in the exclusive set-up (the scattered electron as well as all
component of pions momenta are registered) have a form
\ba
A_c=\frac{2\Sigma Re(\Phi_\rho \Phi_\sigma^*)}{\Sigma|\Phi_\rho|^2\Lambda^2g^2_{\rho\pi\pi}
+\Sigma|\Phi_\sigma|^2g^2_{\sigma\pi\pi}}\Lambda g_{\rho \pi \pi}g_{\sigma \pi \pi},
\ea
where the sum over the fermion's spin states is implied. The
asymmetry as well as the cross section do not depend on $s$ in the
large $s$ limit.
The notations are:
\begin{gather}
\label{desig}
a=\frac{\Lambda \,g\, m\,(1-x)}{2(m^2+Q^2)(q_1^2-m_\rho^2)},\quad
b=-\frac{\Lambda\, g\, m\, qq_1}{2(m^2+Q^2)(q_1^2-m_\rho^2)},\\ \nonumber
c=\frac{g \,m}{2 \Lambda (m^2+Q^2)(q_1^2-m_\sigma^2)}[(qq_--qq_+)(1-x)-(x_--x_+)(qq_++qq_-)],\\ \nonumber
d=(p+q)^2-m_p^2=-Q^2+A_-+A_++A-m_p^2,\\ \nonumber
d'=(p'-q)^2-m_p^2=-Q^2+2\vec{p}\vec{q}-x(A_-+A_++A-m_p^2),
\end{gather}
Here we put the  relevant scalar products, which appear in (\ref{desig}) by using Gribov representation (\ref{g1}-\ref{g3}):
\begin{gather}
\nonumber
p_1^{'2}=p_1^2=0,\quad 2p_1q=-2p_1^{'}q=-Q^2,\quad p^2=(p')^2=m_p^2\, , \\ \nonumber
q^2=-Q^2=(p_1-p_1')^2,\quad p_1q_\pm=\frac{1}{2}sx_\pm, \quad
p_1p=\frac{s}{2},\quad p_1p'=\frac{sx}{2},\\ \nonumber
q_1^2=(p-p')^2=-\frac{1}{x}[m_p^2(1-x)^2+\vec{p}^2], \\ \nonumber
s_1=q_2^2=(q_++q_-)^2=-\frac{1}{x_-x_+}[m_\pi^2(x_-+x_+)^2+(x_+\vec{q}_--x_-\vec{q}_+)^2],\\ \nonumber
qq_1=\frac{1}{2}[s_1+Q^2-q_1^2],\quad
qq_\pm=-\vec{q}\vec{q}_\pm+\frac{x_\pm}{2}(A_-+A_++A-m_p^2),\\ \nonumber
pq_\pm=\frac{1}{2}[m^2_\pi+\vec{q}_\pm^2+m_p^2x_\pm^2]\frac{1}{x_\pm},\quad
p'q_\pm=\frac{1}{2x_\pm x}[m_p^2x_\pm^2+x^2m_\pi^2+(x_\pm\vec{p}-x\vec{q}_\pm)^2],\\ \nonumber
qp=\frac{1}{2}(A_-+A_++A-m_p^2),\quad qp'=-\vec{q}\vec{p}+\frac{x}{2}(A_-+A_++A-m_p^2),\\ \nonumber
pp'=\frac{1}{2x}[m_p^2(1+x^2)+\vec{p}^2],\quad
q_+q_-=\frac{1}{2x_+x_-}[m_\pi^2(x_+^2+x_-^2)+(x_+\vec{q}_--x_-\vec{q}_+)^2].
\end{gather}
\newpage
\begin{center}
\begin{tabular}{||c||c|c|c|c|c|c|c|c||}  \hline\hline
$x_{-}\backslash x_{+}$ & 0.1&0.2&0.3&0.4&0.5&0.6&0.7&0.8 \\ \hline\hline
0.1 &0&-0.32&-0.51&-0.46&-0.37&-0.26&-0.15&-0.07 \\ \hline
0.2 & 0.32&0&0.032&-0.037&-0.071&-0.067&-0.058  & \\ \hline
0.3 & 0.51&-0.03&0&0.001&-0.039&-0.055 & & \\ \hline
0.4 & 0.46&0.037&-0.001&0&-0.019 & & & \\ \hline
0.5 & 0.367&0.071&0.0398&0.0197 & & & & \\ \hline
0.6 & 0.257&0.067&0.055 & & & & & \\ \hline
0.7 & 0.148&0.058 & & & & & & \\ \hline
0.8 & 0.07 & & & & & & & \\ \hline\hline
\end{tabular}
\end{center}
Table 1: Integrated Asymmetry for equal masses of $\sigma$ and $\rho$ mesons,
$Q=1.2$ GeV (for other numerical parameters see Section \ref{Disc}).

\vspace*{0.2cm}
\begin{center}
\begin{tabular}{||c||c|c|c|c|c|c|c|c||}\hline \hline
$x_{-}\backslash x_{+}$ &0.1&0.2&0.3&0.4&0.5&0.6&0.7&0.8 \\ \hline\hline
0.1 &0&-0.09&-0.39&-0.65&-0.81&-0.87&-0.84&-0.71 \\ \hline
0.2 &0.09&0&-0.06&-0.09&-0.32&-0.55&-0.59 & \\ \hline
0.3 &0.39&0.06&0&-0.039&-0.031&-0.065 & & \\ \hline
0.4 &0.653&0.093&0.039&0&-0.021 & & & \\ \hline
0.5 &0.806&0.323&0.031&0.021 & & & & \\ \hline
0.6 &0.869&0.547&0.065 & & & & & \\ \hline
0.7 &0.845&0.595 & & & & & & \\ \hline
0.8 &0.71 & & & & & & & \\ \hline\hline
\end{tabular}
\end{center}
Table 2: Integrated Asymmetry for mass of $\sigma$ equals $1.2$ Gev,
$Q=0.7$ GeV.

\vspace*{0.2cm}
\begin{center}
\begin{tabular}{||c||c|c|c|c|c|c|c|c||} \hline\hline
$x_{-}\backslash x_{+}$ &0.1&0.2&0.3&0.4&0.5&0.6&0.7&0.8 \\ \hline\hline
0.1 &0&-0.83&-0.88&-0.85&-0.79&-0.71&-0.62&-0.48 \\ \hline
0.2 &0.83&0&-0.45&-0.56&-0.54&-0.47&-0.38 & \\ \hline
0.3 &0.88&0.44&0&-0.22&-0.29&-0.28 & & \\ \hline
0.4 &0.85&0.55&0.22&0&-0.11 & & & \\ \hline
0.5 &0.79&0.54&0.29&0.11 & & & & \\ \hline
0.6 &0.72&0.47&0.28 & & & & & \\ \hline
0.7 &0.62&0.38 & & & & & & \\ \hline
0.8 &0.48 & & & & & & & \\ \hline\hline
\end{tabular}
\end{center}

Table 3: Integrated Asymmetry for equal masses of $\sigma$ and $\rho$ mesons,
$Q=0.7$ GeV.

\vspace*{0.2cm}
\begin{center}
\begin{tabular}{||c||c|c|c|c|c|c|c|c||} \hline\hline
$x_{-}\backslash x_{+}$ &0.1&0.2&0.3&0.4&0.5&0.6&0.7&0.8 \\ \hline\hline
0.1 &0&-0.788&-0.938&-0.947&-0.911&-0.841&-0.735&-0.576 \\ \hline
0.2 &0.788&0&-0.545&-0.777&-0.743&-0.642&-0.510 & \\ \hline
0.3 &0.938&0.545&0&-0.299&-0.454&-0.442 & & \\ \hline
0.4 &0.947&0.777&0.299&0&-0.168 & & & \\ \hline
0.5 &0.911&0.743&0.454&0.168 & & & & \\ \hline
0.6 &0.84&0.64&0.44 & & & & & \\ \hline
0.7 &0.73&0.51 & & & & & & \\ \hline
0.8 &0.57 & & & & & & & \\ \hline\hline
\end{tabular}
\end{center}

Table 4: Integrated Asymmetry for mass of $\sigma$ equals $0.469$ Gev,
$Q=0.7$ GeV.


\begin{center}
\begin{tabular}{||c||c|c|c|c|c|c|c|c|c|c|c||} \hline\hline
$\theta_{+} \backslash \theta_{-}$ &0.22&0.18&0.13&0.09&0.04&0&-0.04&-0.09&-0.13&-0.18&-0.22 \\ \hline\hline
0.22&0&0.02&0.06&0.16&0.12&-0.53&0.12&0.16&0.06&0.02&0 \\ \hline
0.18&-0.02&0&0.07&0.23&0.29&-0.34&0.29&0.23&0.07&0&-0.02 \\ \hline
0.13&-0.06&-0.07&0&0.22&0.45&-0.21&0.45&0.22&0&-0.07&-0.06 \\ \hline
0.09&-0.16&-0.23&-0.22&0&0.57&-0.12&0.57&0&-0.22&-0.23&-0.16 \\ \hline
0.04&-0.12&-0.29&-0.45&-0.57&0&0&0&-0.57&-0.45&-0.29&-0.12 \\ \hline
0&0.53&0.34&0.21&0.12&0&0&0&0.12&0.21&0.34&0.53 \\ \hline
-0.04&-0.12&-0.28&-0.45&-0.57&0&0&0&-0.57&-0.45&-0.29&-0.12 \\ \hline
-0.09&-0.16&-0.23&-0.22&0&0.57&-0.12&0.57&0&-0.22&-0.23&-0.16 \\ \hline
-0.13&-0.06&-0.07&0&0.22&0.45&-0.21&0.45&0.22&0&-0.07&-0.06 \\ \hline
-0.18&-0.02&0&0.07&0.23&0.29&0.34&0.29&0.23&0.07&0&-0.02 \\ \hline
-0.22&0&0.02&0.06&0.16&0.12&-0.53&0.12&0.16&0.06&0.02&0 \\ \hline\hline
\end{tabular}
\end{center}
Table 5: Integrated Asymmetry for mass of $\sigma$ equals $0.469$ GeV.

\vspace*{0.2cm}

\begin{center}
\begin{tabular}{||c||c|c|c|c|c|c|c|c|c|c|c||} \hline\hline
$\theta_{+} \backslash \theta_{-}$ &0.22&0.18&0.13&0.09&0.04&0&-0.04&-0.09&-0.13&-0.18&-0.22 \\ \hline\hline
0.22&0&0.03&0.10&0.22&0.09&-0.44&0.09&0.22&0.10&0.03&0 \\ \hline
0.18&-0.03&0&0.13&0.37&0.23&-0.27&0.23&0.37&0.13&0&-0.03 \\ \hline
0.13&-0.10&-0.13&0&0.43&0.34&-0.16&0.34&0.43&0&-0.13&-0.10 \\ \hline
0.09&-0.22&-0.37&-0.43&0&0.36&-0.08&0.36&0&-0.43&-0.37&-0.22 \\ \hline
0.04&-0.09&-0.23&-0.34&-0.36&0&-0.03&0&-0.36&-0.34&-0.23&-0.09 \\ \hline
0&0.43&0.27&0.16&0.08&0.03&0&0.03&0.08&0.16&0.27&0.44 \\ \hline
-0.04&-0.09&-0.23&-0.34&-0.36&0&-0.03&0&-0.36&-0.34&-0.23&-0.09 \\ \hline
-0.09&-0.22&-0.37&-0.43&0&0.36&-0.08&0.36&0&-0.43&-0.37&-0.22 \\ \hline
-0.13&-0.10&-0.13&0&0.43&0.34&-0.16&0.34&0.43&0&-0.13&-0.10 \\ \hline
-0.18&-0.02&0&0.13&0.37&0.23&-0.27&0.23&0.37&0.13&0&-0.03 \\ \hline
-0.22&0&0.03&0.10&0.22&0.09&-0.44&0.09&0.22&0.10&0.03&0 \\ \hline\hline
\end{tabular}
\end{center}
Table 6: Integrated Asymmetry for mass of $\sigma$ equals $0.769$ GeV.

\vspace*{0.2cm}

\begin{center}
\begin{tabular}{||c||c|c|c|c|c|c|c|c|c|c|c||} \hline\hline
$\theta_{+} \backslash \theta_{-}$ &0.22&0.18&0.13&0.09&0.04&0&-0.04&-0.09&-0.13&-0.18&-0.22 \\ \hline\hline
0.22&0&0.04&0.15&0.27&0.05&-0.33&0.05&0.27&0.15&0.04&0\\ \hline
0.18&-0.04&0&0.22&0.44&0.14&-0.19&0.14&0.44&0.22&0&-0.04 \\ \hline
0.13&-0.15&-0.22&0&0.47&0.17&-0.10&0.17&0.47&0&-0.22&-0.15 \\ \hline
0.09&-0.27&-0.44&0.47&0&0.13&-0.03&0.13&0&-0.47&-0.44&-0.27 \\ \hline
0.04&-0.05&-0.14&-0.17&-0.13&0&-0.08&0&-0.13&-0.17&-0.14&-0.05 \\ \hline
0&0.33&0.19&0.10&0.03&0.08&0&0.08&0.03&0.10&0.19&0.33 \\ \hline
-0.04&-0.05&-0.14&-0.17&-0.12&0&-0.08&0&-0.12&-0.17&-0.14&-0.05 \\ \hline
-0.08&-0.27&-0.44&-0.47&0&0.12&-0.03&0.12&0&-0.47&-0.44&-0.27 \\ \hline
-0.13&-0.15&-0.22&0&0.47&0.17&-0.10&0.17&0.47&0&-0.22&-0.15 \\ \hline
-0.18&-0.41&0&0.22&0.44&0.14&-0.19&0.14&0.44&0.22&0&-0.04 \\ \hline
-0.22&0&0.04&0.15&0.27&0.05&-0.33&0.05&0.27&0.15&0.04&0 \\ \hline\hline
\end{tabular}
\end{center}
Table 7: Integrated Asymmetry for mass of $\sigma$ equals $1.200$ GeV.

\end{document}